\documentclass[twocolumn,preprintnumbers,amsmath,amssymb,a4paper]{revtex4-2}
\usepackage{bm}
\usepackage{extarrows}
\usepackage{amsfonts}
\usepackage{amsmath}
\usepackage{amssymb}
\usepackage{geometry}
\usepackage{graphicx}
\usepackage{footmisc}
\usepackage{braket}
\usepackage{array}
\usepackage{lipsum}
\usepackage{cancel}
\usepackage{bibentry,natbib}
\usepackage[colorlinks,linkcolor=blue,anchorcolor=blue,citecolor=blue,urlcolor=blue]{hyperref}

\begin{document}

\title{Revised $^3$He nuclear charge radius due to electronic hyperfine mixing}

\author{Xiao-Qiu Qi$^{1,2}$} 
\author{Pei-Pei Zhang$^{2}$} 
\author{Zong-Chao Yan$^{3,2}$}
\author{Li-Yan Tang$^{2}$}
\author{Ai-Xi Chen$^{1}$}
\author{Ting-Yun Shi$^{2,*}$}
\author{Zhen-Xiang Zhong$^{4,2,\dagger}$}

\affiliation {$^1$ Department of Physics, Zhejiang Sci-Tech University, Hangzhou 310018, China}
\affiliation {$^2$ State Key Laboratory of Magnetic Resonance and Atomic and Molecular Physics, Wuhan Institute of Physics and Mathematics, Innovation Academy for Precision Measurement Science and Technology, Chinese Academy of Sciences, Wuhan 430071, China}
\affiliation {$^3$ Department of Physics, University of New Brunswick, Fredericton, New Brunswick, Canada E3B 5A3}
\affiliation {$^4$ Center for Theoretical Physics, School of Physics and Optoelectronic Engineering, Hainan University, Haikou 570228, China}

\date{\today}

\begin{abstract}
The significant discrepancy in the difference of squared nuclear charge radii $\Delta R^2$ of $^{3,4}$He obtained from electronic-atom or muonic-atom energy levels is a puzzle. In this paper, we show that the tension is resolved by including off-diagonal mixing effects due to the hyperfine interaction. Our findings indicate that the hyperfine mixing effect from the $n\,^3\!S$ and $n\,^1\!S$ states ($n>2$) of $^3$He leads to a $-1.37$ kHz adjustment in the isotope shift of the $2\,^1\!S-2\,^3\!S$ transition, surpassing the current uncertainty by a factor of $7$. This results in a change of $-0.0064~\rm{fm}^2$ in $\Delta R^2$, shifting from $1.0757(15)~\mathrm{fm}^2$ to $1.0693(15)~\mathrm{fm}^2$ as determined by Werf {\it et al.}, significantly reducing the discrepancy with the value of $1.0636(31)~\mathrm{fm}^2$ determined by $\mu\rm{He}^+$, and aligning with the result of $1.069(3)$ $\mathrm{fm}^2$ obtained from the $2\,^3\!S-2\,^3\!P$ transition. This adjustment will result in a noticeable change in the absolute nuclear charge radius of $^{3}$He by $-0.0017~\rm{fm}$, aligning the revised value of $1.9715(11)~\mathrm{fm}$ with the value of $1.97007(94)~\mathrm{fm}$ determined by $\mu^3\rm{He}^+$ within $1\sigma$. Our results offer crucial insights into resolving discrepancy in $\Delta R^2$ for $^{3,4}$He and determining the charge radius of $^3$He.
\end{abstract}

\maketitle

\emph{Introduction.}---
The proton size puzzle has attracted a tremendous amount of attention in the physics community (see Refs.~\cite{Gao2022,Scheidegger2024,Brandt2022} and references therein) 
and has stimulated experimental searches for similar anomalies in other light atomic nuclei such as D, $^3$He, and $^4$He~\cite{Pohl2016,Werf2023,Rengelink2018,Schuhman2023,Krauth2021}. The puzzle arises from an apparent discrepancy between the values obtained from electronic energy levels or scattering and those derived from muonic energy levels. For instance, the nuclear charge radius of $^3$He is measured as $1.9732(11)$~fm from electronic $^3$He energy levels~\cite{Werf2023,Rengelink2018}, while the value obtained from $\mu^3$He$^+$~\cite{Schuhman2023} is $1.97007(94)$~fm, showing a lack of agreement.

At present, precise spectroscopic measurements of helium and helium-like ions are among the primary methods for acquiring information about nuclear structure~\cite{Pastor2004,Pastor2012,Qi2020,Guan2020,Qi2023,Sun2023,Drake2013}. This includes measurements of the nuclear charge radius, Zemach radius, electric quadrupole moment, and other related properties.
However, while the current experiments have achieved impressive accuracy in measuring the nuclear charge radii of $^3$He and $^4$He individually~\cite{Werf2023,Rengelink2018,vanRooij2011}, the precision of the theoretical energy levels remains inadequate~\cite{Yan1995,Pachucki2006A,Pachucki2006X,Yerokhin2010,Pachucki2019,Patkos2021a,Patkos2021b,Patkos2023}. Consequently, calculating the isotope shift of a transition frequency has become an effective method for precisely determining the difference in nuclear charge radii between these two isotopes, as it can cancel out nuclear mass-independent contributions and their associated uncertainties.

A comprehensive analysis of the isotope shifts and the nuclear charge radii difference $\Delta R^2$ between $^3$He and $^4$He was conducted by Pachucki {\it et al.}~\cite{Pachucki2015}. 
In 2016 and 2017, higher-order recoil corrections of order $\alpha^6 m^2/M$ for the triplet and singlet states of the helium atom were completed~\cite{Patkos2016,Patkos2017}, and the values of 
$\Delta R^2$ were subsequently updated to $1.061(3)~\mathrm{fm}^2$ from Ref.~\cite{Shiner1995} and $1.069(3)~\mathrm{fm}^2$ from Ref.~\cite{Pastor2012} for the $2\,^3\!S-2\,^3\!P$ transition, and $1.027(11)~\mathrm{fm}^2$ from Ref.~\cite{vanRooij2011} for the $2\,^1\!S-2\,^3\!S$ transition. 
There is an unexpected and clear discrepancy in the extracted values of $\Delta R^2$ from these two electronic transitions.

Aiming at resolving the unexplained discrepancy in $\Delta R^2$ of $^{3,4}$He, numerous high-precision spectroscopic measurements have been undertaken in recent years~\cite{Zheng2017,Rengelink2018,Huang2020,Werf2023,Krauth2021,Schuhman2023}.
Zheng {\it et al.}~\cite{Zheng2017} updated $\Delta R^2$ to $1.028(2)~\mathrm{fm}^2$ by measuring the $2\,^3\!S-2\,^3\!P$ transition frequency. Rengelink {\it et al.}~\cite{Rengelink2018} and Werf {\it et al.}~\cite{Werf2023} measured the $2\,^1\!S-2\,^3\!S$ transition frequency, resulting in updated $\Delta R^2$ of $1.041(7)~\mathrm{fm}^2$ and $1.0757(15)~\mathrm{fm}^2$, respectively. Huang {\it et al.}~\cite{Huang2020} performed laser spectroscopy of the $2\,^1\!S_0-3\,^1\!D_2$ two-photon transition in $^3$He and determined $\Delta R^2$ to be
1.059(25)~fm$^2$. 
Very recently, the nuclear charge radii for $^3$He and $^4$He have been precisely extracted as $1.97007(94)$~fm~\cite{Schuhman2023} and $1.6786(12)$~fm~\cite{Krauth2021}, respectively, by measuring the $2S-2P$ transition in $\mu^3\mathrm{He}^+$ and $\mu^4\mathrm{He}^+$, yielding a value of $1.0636(31)~\mathrm{fm}^2$ for $\Delta R^2$.

In addition, Ottermann {\it et al.}~\cite{Ottermann1985} determined the charge radii to be $1.671(14)~\mathrm{fm}$ for $^4$He and $1.976(15)~\mathrm{fm}$ for $^3$He from electron scattering experiments, resulting in a corresponding $\Delta R^2$ of $1.112(75)~\mathrm{fm}^2$. 
A comprehensive study of global data on elastic electron-helium scattering by Sick~\cite{Sick2008} has yielded a more accurate determination of the charge radius of  $^4$He to be $1.681(4)~\mathrm{fm}$, and thus updated the value of $\Delta R^2$ to $1.079(61)~\mathrm{fm}^2$. 
Later, Sick~\cite{Sick2014} further revised the value of $\Delta R^2$ to $1.066(60)~\mathrm{fm}^2$. However, there have been no substantial developments in improving the accuracy of electron-$^3$He scattering measurements. 

Therefore, the perplexing discrepancy still exists between the $\Delta R^2$ values derived from electronic-atom and muonic-atom spectra. Additionally, the accuracy of electron scattering measurements remains insufficient. As a result, the nuclear charge radius of $^3$He cannot yet be determined with precision. This paper aims to reassess the situation by examining the impact of fine and hyperfine mixing effects on the isotope shifts of the $2\,^1\!S_0-2\,^3\!S_1$ and $2\,^3\!S_1-2\,^3\!P_J$ transitions between $^3$He and $^4$He. By incorporating the hyperfine mixing effect, we have nearly resolved the discrepancy and successfully reconciled the $^3$He nuclear charge radius obtained from different methods.

\begin{table*}[htbp!]
	\caption{\label{tab:IS} Isotope shifts between $^3$He and $^4$He, considering point-like nuclei, in kHz. The anticipated numerical uncertainty is typically less than 1~kHz for the last significant digit, unless otherwise specified.}
	\begin{center}
		\begin{tabular}{l r@{}l r@{}l c r@{}l r@{}l}
			\hline
			\hline
			\multicolumn{1}{c}{ }
			&\multicolumn{4}{c}{$2^1S-2^3S$}
			&\multicolumn{1}{c}{ }
			&\multicolumn{4}{c}{$2^3S-2^3P$} \\
			\cline{2-5} \cline{7-10}
			\multicolumn{1}{c}{ }
			&\multicolumn{2}{c}{This work}
			&\multicolumn{2}{c}{Pachucki  {\it et al.}~\cite{Pachucki2015}} 
			&\multicolumn{1}{c}{ }
			&\multicolumn{2}{c}{This work}
			&\multicolumn{2}{c}{Pachucki  {\it et al.}~\cite{Pachucki2015}} \\
			\hline
			$\mu\alpha^2$                                                                               &$-8632567$&$.86      $ &$-8632567$&$.86      $ & &$12412458$&$.1       $ &$12412458$&$.1       $ \\
			$\mu\alpha^2(\mu/M)$                                                                        &$  608175$&$.58      $ &$  608175$&$.58      $ & &$21243041$&$.3       $ &$21243041$&$.3       $ \\
			$\mu\alpha^2(\mu/M)^2$                                                                      &$   -7319$&$.80      $ &$   -7319$&$.80      $ & &$   13874$&$.6       $ &$   13874$&$.6       $ \\
			$\mu\alpha^2(\mu/M)^3$                                                                      &$       0$&$.30      $ &$       0$&$.30      $ & &$       4$&$.6       $ &$       4$&$.6       $ \\
			$\mu\alpha^4$                                                                               &$   -8954$&$.22      $ &$   -8954$&$.22      $ & &$   17872$&$.8       $ &$   17872$&$.8       $ \\
			$\mu\alpha^4(\mu/M)$                                                                        &$    6458$&$.22      $ &$    6458$&$.23      $ & &$  -20082$&$.3       $ &$  -20082$&$.4       $ \\
			$\mu\alpha^4(\mu/M)^2$                                                                      &$       1$&$.84      $ &$       1$&$.84      $ & &$      -3$&$.0       $ &$      -3$&$.0       $ \\
			$m\alpha^5(m/M)$                                                                            &$      56$&$.59      $ &$      56$&$.61      $ & &$     -60$&$.6       $ &$     -60$&$.7       $ \\
			$\sim m\alpha^6(m/M)$                                                                       &$       2$&$.33(59)  $ &$       2$&$.75(69)  $ & &$     -13$&$.1(3.3)  $ &$     -15$&$.5(3.9)  $ \\
			$E^{n=2}_{\rm mix,hfs}$                                                                     &$      80$&$.76      $ &$      80$&$.72      $ & &$      54$&$.6       $ &$      54$&$.6       $ \\
			Nuclear polarization                                                                                        &$       0$&$.20(2)   $ &$       0$&$.20(2)   $ & &$      -1$&$.1(1)    $ &$      -1$&$.1(1)    $ \\
			Sum                                                                                         &$-8034066$&$.06(59)  $ &$-8034065$&$.66(69)  $ & &$33667145$&$.9(3.3)  $ &$33667143$&$.2(3.9)  $ \\
			$m\alpha^6(m/M)$                                                                            &$       2$&$.73^a    $ &$        $&$         $ & &$      -9$&$.4^a     $ &$        $&$         $ \\
			$\sim m\alpha^6(m/M)^2$                                                                     &$       0$&$.00(15)^a$ &$        $&$         $ & &$        $&$         $ &$        $&$         $ \\
			$\sim m\alpha^7(m/M)$                                                                       &$      -0$&$.21(11)^a$ &$        $&$         $ & &$       0$&$.0(0.9)^a$ &$        $&$         $ \\
			Isotope shifts                                                                              &$-8034065$&$.87(19)  $ &$-8034065$&$.91(19)^a$ & &$33667149$&$.6(0.9)  $ &$33667149$&$.3(0.9)^a$ \\    
			$\delta E_{\rm mix,hfs}=E_{\mathrm{mix},\mathrm{hfs}}-E^{n=2}_{\mathrm{mix},\mathrm{hfs}}$  &$      -1$&$.37      $ &$        $&$         $ & &$       0$&$.09      $ &$        $&$         $ \\
			Fine structure mixing                                                                                   &$        $&$         $ &$        $&$         $ & &$      -0$&$.43      $ &$        $&$         $ \\
			Isotope shifts (This work)                                                                  &$-8034067$&$.24(19)  $ &$        $&$         $ & &$33667149$&$.3(0.9)  $ &$        $&$         $ \\
			Ref.~\cite{Morton2006}                                                                      &$-8034067$&$.8(1.1)  $ &$        $&$         $ & &$33667146$&$.2(7)    $ &$        $&$         $ \\      
			\hline
			\hline
		\end{tabular}
		\begin{minipage}{0.9\textwidth}
			\begin{flushleft}
			$^a$ The values are cited from Refs.~\cite{Patkos2016,Patkos2017}. \\
			$\sim$ Values obtained through approximate evaluation.
			\end{flushleft}
		\end{minipage}
	\end{center}
\end{table*}

\emph{Fine and hyperfine mixing.}---
The fine and hyperfine mixing effects can influence isotope shifts and fine structure splittings, typically on the order of $m\alpha^6$. Accurately computing these effects has always been a challenging task.
Currently, the primary approach approximates the contributions of fine and hyperfine mixing effects by focusing on the dominant states, such as the mixed effects of $2\,^3\!S-2\,^1\!S$ and $2\,^1\!P-2\,^3\!P$. However, as experimental measurement accuracy continues to improve, these approximations may no longer suffice to meet the demands for theoretical precision. Additionally, it is important to consider whether the correction of higher-order perturbations has affected the current level of accuracy.
To address this, we will compute the second-order perturbation and evaluate the necessity of incorporating higher-order contributions through a comparative analysis of perturbation theory (Method 1) and exact diagonalization (Method 2).

Here, we use the $2\,^1\!P$ and $2\,^3\!P$ states as an example to introduce two distinct methods for studying this effect.
Method 1 involves performing an exact diagonalization within the $2\,^3\!P$ manifold while addressing the $2\,^1\!P-2\,^3\!P$ mixing effect using second-order perturbation theory, a technique also employed by Pachucki {\it et al.}~\cite{Pachucki2015}.
Method 2 expands the $2\,^3\!P$ manifold to include the $2\,^1\!P$ state, followed by an exact diagonalization of the extended matrix.

In Method 1, the contribution of mixing effects consists of two components: fine structure (fs) and hyperfine structure (hfs):
\begin{equation}\label{eq:1}
\begin{aligned}
E_{\mathrm{mix}} = E_{\mathrm{mix},\mathrm{fs}} +E_{\mathrm{mix},\mathrm{hfs}},
\end{aligned}
\end{equation}
where
\begin{equation}\label{eq:2}
\begin{aligned}
E_{\mathrm{mix},\mathrm{fs}} (2\,^3\!P) = \sum_{n} \frac{ \vert \langle 2\,^3\!\vec{P} \vert H_{\mathrm{fs}}^{(4)} \vert n\,^1\!\vec{P} \rangle \vert^2}{E(2\,^3\!P)-E(n\,^1\!P)} ,
\end{aligned}
\end{equation}
and
\begin{equation}\label{eq:3}
\begin{aligned}
E_{\mathrm{mix},\mathrm{hfs}} (2\,^3\!\chi) = \sum_{n} \frac{\vert \langle 2\,^3\!\vec{\chi} \vert H_{\mathrm{hfs}}^{(4)} \vert n\,^1\!\vec{\chi} \rangle \vert^2}{E(2\,^3\!\chi)-E(n\,^1\!\chi)},\; \chi\in \{S,P\}\,.
\end{aligned}
\end{equation}
In the above, the initial state is on the left, while the intermediate state is on the right. Here, $H_{\mathrm{fs}}^{(4)}$ represents the spin-dependent component of Eq.~(17) in the Supplemental Material~\cite{SM}, and $H_{\mathrm{hfs}}^{(4)}$ accounts for the leading-order hyperfine structure splitting~\cite{Pachucki2012,Pachucki2015}. Following the methodology outlined in Ref.~\cite{Pachucki2012}, the relevant formulas for calculating the impact of hfs mixing on the isotope shift are as follows:
\begin{equation}\label{eq:4}
    \begin{aligned}
        E_{\mathrm{mix},\mathrm{hfs}}(2\,^3\!S) =& C^2_{56} \cdot \frac{1}{4} \sum_{n} \frac{\vert\langle 2\,^3\!S \vert Q_A \vert n^1\!S \rangle\vert^2}{E(2\,^3\!S)-E(n^1\!S)} ,
    \end{aligned}
\end{equation}
\begin{equation}\label{eq:5}
    \begin{aligned}
        E_{\mathrm{mix},\mathrm{hfs}}(2^1\!S) =& C^2_{56} \cdot \frac{3}{4} \sum_{n} \frac{|\langle 2^1\!S \vert Q_A \vert n\,^3\!S \rangle|^2}{E(2^1\!S)-E(n\,^3\!S)},
    \end{aligned}
\end{equation}
and
\begin{equation}\label{eq:6}
    \begin{aligned}
&E_{\mathrm{mix},\mathrm{hfs}}(2\,^3\!P)= C^2_{56} \cdot \sum_{n} \frac{1}{E(2\,^3\!P)-E(n^1\!P)} \\
&\cdot \bigg[ \frac{1}{4} |\langle 2\,^3\!P \vert Q_A \vert n^1\!P \rangle|^2 + \frac{1}{20} |\langle 2\,^3\!P \vert \hat{Q}_A \vert n^1\!P \rangle|^2 \bigg],
    \end{aligned}
\end{equation}
where
\begin{equation}\label{eq:7}
\begin{aligned}
Q_A= \frac{4\pi Z}{3}[\delta^3(r_1)-\delta^3(r_2)],
\end{aligned}
\end{equation}
and
\begin{equation}\label{eq:8}
\begin{aligned}
\hat{Q}_A= \frac{Z}{2} \bigg[\frac{1}{r_1^3} \bigg( \delta^{ij} - 3\frac{r_1^ir_1^j}{r_1^2} \bigg)-\frac{1}{r_2^3} \bigg( \delta^{ij} - 3\frac{r_2^ir_2^j}{r_2^2} \bigg) \bigg].
\end{aligned}
\end{equation}
The coefficient $C^y_{xz}$ here is defined as $\mu^x [(1+\kappa)/(mM)]^y \alpha^z$,
where $\kappa$ represents the anomalous magnetic moment of the nucleus, $M$ and $ m $ are the masses of the nucleus and the electron, respectively, $\mu$ is the reduced mass, and $Z$ denotes the nuclear charge. It is important to note that the second-order perturbation calculation of the delta function does not converge, which necessitates the process of renormalization. For details, refer to 
Ref.~\cite{Pachucki2012} and the Supplementary Material \cite{SM}.
Furthermore, since the contribution of the $\hat{Q}^A$ term in the second part of Eq.~(\ref{eq:6}) is negligible, we consider only the $2\,^1\!P-2\,^3\!P$ mixing within this term.

In Method 2, the contribution of the mixing effects is determined by diagonalizing the following matrix:
\begin{equation}\label{eq:9}
\begin{aligned}
\begin{bmatrix}
E(n\,^1\!\chi_J)                &                   &        &                                                    &                 \\
                                &E(n\,^3\!\chi_J)   &        &\multicolumn{2}{c}{\langle \mathcal{H} \rangle_J^F}                   \\
                                &                   &\ddots  &                                                    &                 \\
\multicolumn{2}{c}{\langle \mathcal{H} \rangle_J^F} &        &E(2\,^1\!\chi_J)                                    &                 \\
                                &                   &        &                                                    &E(2\,^3\!\chi_J) \\
\end{bmatrix},
\end{aligned}
\end{equation}
where
\begin{equation}\label{eq:10}
\begin{aligned}
\langle \mathcal{H} \rangle_J^F = \langle n\,^1\!\vec{\chi}_J^F \vert H_{\mathrm{fs}}^{(4)}+H_{\mathrm{hfs}}^{(4)} \vert n'\,^3\!\vec{\chi}_J^F \rangle + \rm{higher\;order}.
\end{aligned}
\end{equation}
The diagonal component represents the fine structure energy relative to the $2\,^3\!\chi$ (or $2\,^1\!\chi$) centroid, without considering the fine structure mixing. A specific example of this matrix is provided in the Supplemental Material~\cite{SM}. Next, calculate the correction for each energy level, $\delta E(2\,^1\!\chi_J^F)$ and $\delta E(2\,^3\!\chi_J^F)$, which leads to the determination of the associated contribution to the isotope shift, denoted as
\begin{equation}\label{eq:11}
\begin{aligned}
E_{\mathrm{mix}} = \frac{\sum_{J,F} (2F+1) \delta E(2\,^{2S+1}\!\chi_J^F)}{(2I+1)(2S+1)(2L+1)},
\end{aligned}
\end{equation}
where the $L$ and $S$ are, respectively, the orbital and spin angular momenta of the state, and $I$ is the nuclear spin.

\begin{table}[htbp!]
	\caption{Assessing the hyperfine mixing effect on the $^3\rm{He}-^4\rm{He}$ isotope shifts for the $2^1\!S$, $2^3\!S$, and $2^3\!P$ states, in kHz. Here, M1 and M2 denote Method 1 and Method 2, respectively.
	}\label{tab:IS2}
	\begin{tabular}{l r@{}l r@{}l c r@{}l r@{}l c r@{}l r@{}l}
		\hline
		\hline
		\multicolumn{1}{c}{} 
		&\multicolumn{4}{l}{$n\,^3\!S-2\,^1\!S$}
		&\multicolumn{1}{c}{} 
		&\multicolumn{4}{c}{$n\,^1\!S-2\,^3\!S$}
		&\multicolumn{1}{c}{} 
		&\multicolumn{4}{c}{$n\,^1\!P-2\,^3\!P$}\\
		\cline{2-5}\cline{7-10}\cline{12-15}
		\multicolumn{1}{c}{} 
		&\multicolumn{2}{c}{M1}
		&\multicolumn{2}{c}{M2}
		&\multicolumn{1}{c}{} 
		&\multicolumn{2}{c}{M1}
		&\multicolumn{2}{c}{M2}
		&\multicolumn{1}{c}{} 
		&\multicolumn{2}{c}{M1}
		&\multicolumn{2}{c}{M2}\\
		\hline
		$E^{n=2}_{\rm mix,hfs}$ &$60$&$.57$ &$60$&$.57$ & &$-20$&$.19$ &$-20$&$.19$ & &$-74$&$.80$ &$-74$&$.73$ \\
		$E^{n=3}_{\rm mix,hfs}$ &$-1$&$.12$ &$-1$&$.12$ & &$ -0$&$.07$ &$ -0$&$.07$ & &$ -0$&$.04$ &$ -0$&$.04$ \\
		$E^{n=4}_{\rm mix,hfs}$ &$-0$&$.17$ &           & &$   $&$   $ &     &      & &$   $&$   $ &     &      \\
		$E^{n=5}_{\rm mix,hfs}$ &$-0$&$.06$ &           & &$   $&$   $ &     &      & &$   $&$   $ &     &      \\
		Sum                     &$59$&$.22$ &    &      & &$-20$&$.26$ &     &      & &$-74$&$.84$ &$   $&$   $ \\
		$E_{\rm mix,hfs}      $ &$59$&$.46$ &    &      & &$-19$&$.93$ &     &      & &$-74$&$.64$ &$   $&$   $ \\
		\hline
		\hline
	\end{tabular}
\end{table}

\emph{Numerical results.}---
This section first calculates isotope shifts using the method adopted by Pachucki {\it et al.}~\cite{Pachucki2015} to independently verify their results for the $2\,^1\!S-2\,^3\!S$ and $2\,^3\!S-2\,^3\!P$ transitions in $^3\mathrm{He}-^4\mathrm{He}$. We performed variational calculations using the Hylleraas basis set~\cite{Qi2020,Qi2023}, incorporating various relativistic and QED corrections (see Supplemental Material~\cite{SM}), and employed Method 1 to assess the $2\,^1\!S-2\,^3\!S$ and $2\,^1\!P-2\,^3\!P$ mixing effects.
The nuclear parameters used are as follows: the nuclear masses are $M{(^3\mathrm{He})} = 5495.88528007(24)\, m_e$ for $^3$He and $M{(^4\mathrm{He})} = 7294.29954142(24)\, m_e$ for $^4$He, and the nuclear anomalous magnetic moment of $^3$He is $\kappa = -4.18415368(3)$~\cite{CODATA2018}.
Our results are summarized in Table~\ref{tab:IS}.

\begin{table*}[htbp!]
  \caption{\label{tab:He3} Difference in squared nuclear charge radii between $^3\rm{He}$ and $^4\rm{He}$ (in fm$^2$), along with the nuclear charge radius of $^3\rm{He}$ (in fm). The corrected values are defined as $\Delta R^2 = \Delta R_0^2 + \delta \Delta R^2$ and $R({^3\rm{He}}) = R_0({^3\rm{He}}) + \delta R({^3\rm{He}})$, where $\delta \Delta R^2 = -0.0064$~$\rm{fm}^2$ and $\delta R({^3\rm{He}}) = -0.0017$~$\rm{fm}$. The nuclear charge radius of $^4\rm{He}$ is $R({^4\rm{He}}) = 1.6786(12)$~$\rm{fm}$, as reported in Ref.~\cite{Schuhman2023}.
}
  \begin{ruledtabular}
  \begin{tabular}{llllll}
   \multicolumn{1}{l}{} 
  &\multicolumn{1}{c}{$\Delta R_0^2$}
  &\multicolumn{1}{c}{$\Delta R^2$}
  &\multicolumn{1}{c}{$R_0({^3\rm{He}})$}
  &\multicolumn{1}{c}{$R({^3\rm{He}})$}
  &\multicolumn{1}{c}{Ref.} \\
  \hline
  $2\,^1\!S-2\,^3\!S             $ &$1.027(11) $ &$1.021(11) $ &$1.9608(30)$ &$1.9591(30) $ &Theo.~\cite{Pachucki2015,Patkos2016,Patkos2017};Expt.~\cite{vanRooij2011}               \\
  $                              $ &$1.041(7)  $ &$1.035(7)  $ &$1.9644(21)$ &$1.9627(21) $ &Theo.~\cite{Pachucki2015,Patkos2016,Patkos2017};Expt.~\cite{vanRooij2011,Rengelink2018} \\
  $                              $ &$1.0757(15)$ &$1.0693(15)$ &$1.9732(11)$ &$1.9715(11) $ &Theo.~\cite{Pachucki2015,Patkos2016,Patkos2017};Expt.~\cite{Werf2023,Rengelink2018}     \\
  $2\,^3\!S-2\,^3\!P             $ &$          $ &$1.069(3)  $ &$          $ &$1.9715(13) $ &Theo.~\cite{Pachucki2015,Patkos2016,Patkos2017};Expt.~\cite{Pastor2004,Pastor2012}      \\
  $                              $ &$          $ &$1.061(3)  $ &$          $ &$1.9694(13) $ &Theo.~\cite{Pachucki2015,Patkos2016,Patkos2017};Expt.~\cite{Shiner1995}                 \\
  $                              $ &$          $ &$1.028(2)  $ &$          $ &$1.9610(11) $ &Theo.~\cite{Pachucki2015,Patkos2016,Patkos2017};Expt.~\cite{Pastor2012,Zheng2017}       \\
  $2\,^1\!S-3\,^1\!D             $ &$          $ &$1.059(25) $ &$          $ &$1.9689(64) $ &Theo.~\cite{Morton2006};Expt.~\cite{Huang2020}                                          \\  
  $\mu\,^{3,4}\mathrm{He}^+:2S-2P$ &$          $ &$1.0636(31)$ &$          $ &$1.97007(94)$ &Theo.~\cite{Pachucki2023};Expt.~\cite{Krauth2021,Schuhman2023}                          \\
  Electron scattering (ES)         &$          $ &$1.066(60) $ &$          $ &$1.973(14)  $ &Expt.~\cite{Ottermann1985,Sick2008,Sick2014}                                            \\
  \end{tabular}
  \end{ruledtabular}
\end{table*}

It is noteworthy that our computed values for $\sim m\alpha^6(m/M)$ differ significantly from those of Pachucki {\it et al.}~\cite{Pachucki2015}, primarily due to an incorrect formula used by them, see Eq.~(27) in Ref.~\cite{Pachucki2015}. This error has since been rectified in their subsequent articles on $m\alpha^6(m/M)$ calculations~\cite{Patkos2016,Patkos2017}. 
Since the evaluation of this part is difficult, we opted for the methodology employed in Ref.~\cite{Pachucki2015}, {\it i.e.}, we approximated the calculation directly by using the formulas (28)-(31) provided in the Supplemental Material~\cite{SM}, accompanied by an error margin of 25\%.
Furthermore, relying on the comprehensive $m\alpha^6(m/M)$ results and approximate higher-order evaluations presented in the Refs.~\cite{Patkos2016,Patkos2017}, we also list the results of isotope shifts in Table~\ref{tab:IS}, which are consistent with the values of Pachucki {\it et al.}~\cite{Patkos2016,Patkos2017}.

Subsequently, we performed a thorough analysis of the contributions of fine and hyperfine mixing effects using both Method 1 and Method 2. Table~\ref{tab:IS2} outlines the impact of the mixing effects on the isotope shifts of $2^1\!S$, $2^3\!S$, and $2^3\!P$ states.
From the table, it is evident that the contribution of the mixing effect from the $3^3\!S$ and $4^3\!S$ states are significantly greater than that of the $3^1\!S$ and $3^1\!P$ states.
Our findings also indicate that, at the current level of accuracy, there is no noticeable difference between Method 1 and Method 2, suggesting that high-order corrections such as the cross terms between fine and hyperfine structures can be disregarded. However, as future advancements in measurements are made, the exact diagonalization method (Method 2) may become necessary.
Furthermore, by applying the renormalized delta function~\cite{SM}, we also calculated the contributions of the $n\,^3\!S-2\,^1\!S$, $n\,^1\!S-2\,^3\!S$, and $n\,^1\!P-2\,^3\!P$ mixing effects using Method 1, as indicated by $E_{\rm mix,hfs}$ in Table~\ref{tab:IS2}. 
It is noteworthy that the mixing effect originating the $n\,^3\!S$ and $n\,^1\!S$ states significantly impacts the $2\,^1\!S-2\,^3\!S$ transition, leading to a $-1.37$~kHz change in the isotope shift—approximately seven times the uncertainty of the current result, as shown in Table~\ref{tab:IS}.

\begin{figure}[htbp!]
\includegraphics[width=0.5\textwidth]{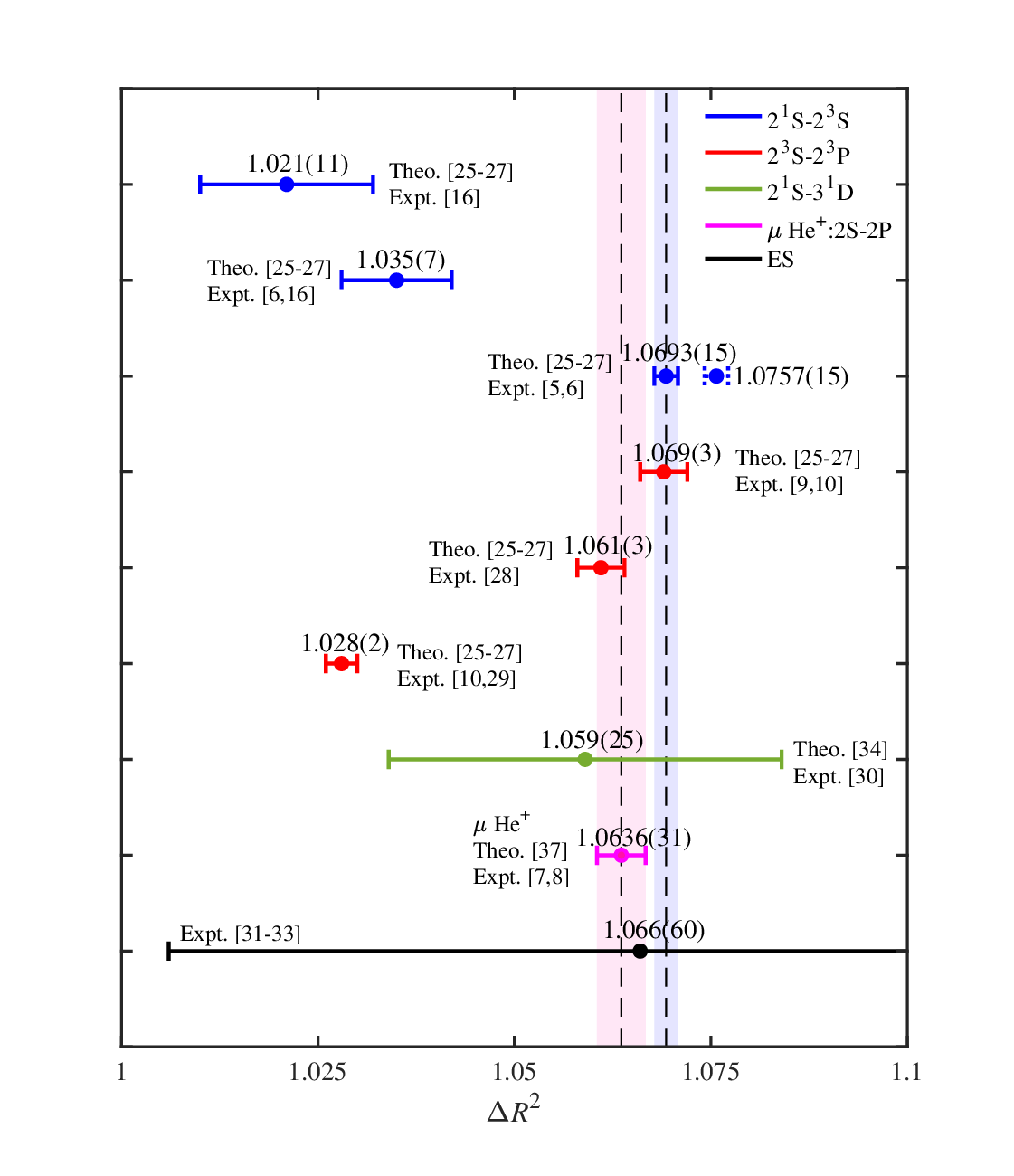}
\caption{Difference of squared nuclear charge radii for $^{3,4}\rm{He}$, in fm$^2$, where similar colors indicate results obtained using the same method. The blue dashed line represents the data before the update, while the blue solid line represents the data after the update. 
}
\label{fig:1}
\end{figure}

Below, we analyze the contributions of $\delta E_{\rm mix,hfs}$ to the difference in squared nuclear charge radius $\Delta R^2$. According to isotope shift theory, the field shift can be expressed as follows:
\begin{equation}\label{eq:12}
\begin{aligned}
E_{\mathrm{FS}}= C \Delta R^2\,,
\end{aligned}
\end{equation}
where the field shift coefficient $C=C_X-C_Y$ for two involved atomic states $X$ and $Y$, and 
\begin{equation}\label{eq:13}
\begin{aligned}
C_{X,Y}= \frac{2\pi Z}{3} \bigg\langle \sum_i \delta^3({r}_i) \bigg\rangle_{X,Y}\,.
\end{aligned}
\end{equation}
The values of coefficient $C$ in Eq.~(\ref{eq:12}) for the $2\,^1\!S-2\,^3\!S$ and $2\,^3\!S-2\,^3\!P$ transitions are calculated to be
\begin{equation}\label{eq:14}
\begin{aligned}
C(2\,^1\!S-2\,^3\!S)=-214.45~\mathrm{kHz/fm}^2\,,
\end{aligned}
\end{equation}
and
\begin{equation}\label{eq:15}
\begin{aligned}
C(2\,^3\!S-2\,^3\!P)=1210.5~\mathrm{kHz/fm}^2\,.
\end{aligned}
\end{equation}

The impact of the $\delta E_{\rm mix,hfs}$ on $\Delta R^2$ can be assessed according to
\begin{equation}\label{eq:16}
	\begin{aligned}
	  \delta \Delta R^2&=- \frac{\delta E_{\rm mix,hfs}}{C(2\,^1\!S-2\,^3\!S)} =-0.0064~\rm{fm}^2\;. \\
	\end{aligned}
\end{equation} 
It can be observed that the inclusion of $\delta E_{\rm mix,hfs}$ causes a change of $-0.0064$~$\rm{fm}^2$ in $\Delta R^2$, as determined by the $2\,^1\!S-2\,^3\!S$ transition. This, in turn, results in a significant alteration of $-0.0017$~$\rm{fm}$ in the absolute nuclear charge radius of $^3$He. 
The change in $\Delta R^2$ determined by the $2\,^3\!S-2\,^3\!P$ transition is only $0.0004$~$\rm{fm}^2$, which falls within the current margin of error, so we will disregard it here.
The updated results for $\Delta R^2$ and $R(^3\rm{He})$ are presented in Table~\ref{tab:He3} and Figure~\ref{fig:1}.

From Table~\ref{tab:He3} and Figure~\ref{fig:1}, we observe that our updated results adjust the difference in squared nuclear charge radii, determined by the latest measured $2\,^1\!S-2\,^3\!S$ transition, from $1.0757(15)~\mathrm{fm}^2$ to $1.0693(15)~\mathrm{fm}^2$. This revision significantly reduces the discrepancy with the value of $1.0636(31)~\mathrm{fm}^2$ obtained from $\mu\rm{He}^+$, narrowing the initial difference from nearly $3\sigma$ to $1.2\sigma$. 
Notably, the revised result also aligns well with the value of $1.069(3)~\mathrm{fm}^2$ determined by the $2\,^3\!S-2\,^3\!P$ transition, sharing the same central value. Additionally, the mixing effect between the $n\,^3\!S$ and $n\,^1\!S$ states results in a change in the nuclear charge radius of $^3\rm{He}$, from $1.9732(11)~\rm{fm}$ to $1.9715(11)~\rm{fm}$, consistent with the value of $1.97007(94)~\rm{fm}$ obtained from $\mu^3\rm{He}^+$.

\emph{Summary.}---
This study investigates the influence of fine and hyperfine mixing effects on isotope shifts, employing both perturbation theory and exact diagonalization methods. Initially, we independently verified the findings of Pachucki {\it  et al.} regarding isotope shifts for the $2\,^1\!S-2\,^3\!S$ and $2\,^3\!S-2\,^3\!P$ transitions in $^{3}\mathrm{He}-^{4}\mathrm{He}$ through numerical calculations. Subsequently, we further examined the effects of mixing in the $n\,^3\!S-2\,^1\!S$, $n\,^1\!S-2\,^3\!S$, and $n\,^1\!P-2\,^3\!P$ transitions.
Our results indicate that there is no significant difference between the perturbation and exact diagonalization methods at the current level of accuracy. Moreover, the hyperfine mixing effect between the $n\,^3\!S$ and $n\,^1\!S$ states has a noteworthy impact on the $2\,^1\!S-2\,^3\!S$ transition, resulting in a considerable shift of $-1.37$ kHz in the isotope shift.
This alternation in the isotope shift for the $2\,^1\!S-2\,^3\!S$ transition also results in a change of $-0.0064$~$\rm{fm}^2$ in the corresponding value of $\Delta R^2$. This adjustment helps to clarify the previously observed discrepancy in the squared difference of charge radii. Meanwhile, our result has also validated that extracting the $\Delta R^2$ through isotope shift is reliable and can be broadly applied to other helium-like systems. Additionally, the revised nuclear charge radius of $^3\rm{He}$ now aligns with the value obtained from measurements of $\mu^3\rm{He}^+$, providing a crucial foundation for accurately assessing the size and structure of the $^3$He nucleus.

In light of these findings, further high-precision measurements of isotope shifts and electron scattering experiments, particularly those involving $^3\rm{He}$, are strongly encouraged to refine the nuclear charge radii of both $^3\rm{He}$ and $^4\rm{He}$, as well as to test nuclear structure models.

The authors extend their gratitude to G. W. F. Drake for his insightful discussions. They also wish to thank Fangfei Wu, Kelin Gao, and Hua Guan for helpful discussions.
This research was supported by the National Natural Science Foundation of China under Grant Nos.\ 12204412, 12393821, 12274423, 12174400, 12175199 and 92265206, and the Science Foundation of Zhejiang Sci-Tech University under Grant No. 21062349-Y. Z.-C.Y acknowledges the support by the Natural Sciences and Engineering Research Council of Canada and by the Digital Research Alliance of Canada. All the calculations were performed on the APM-Theoretical Computing Cluster(APM-TCC).

$*$Email Address: tyshi@wipm.ac.cn

$\dagger$Email Address: zxzhong@hainanu.edu.cn

%

\end{document}